# Strain-Mediated Lattice Reconstruction Enhances Ferromagnetism in $Cr_2Ge_2Te_6$/$WTe_2$ van der Waals Heterobilayers

*Franz Herling[1†\*], Mireia Torres-Sala[1,2†\*], Dorye. L. Esteras[1], Charlotte Evason[1], Motomi Aoki[1], Marcos Rosado[1], Kapil Gupta[1], Bernat Mundet[1], Kai Xu[3], J. Sebastián Reparaz[3], Kenji Watanabe[4], Takashi Taniguchi[5], Dimitre Dimitrov[6,7], Vera Marinova[6], Ivan A. Verzhbitskiy[8], Goki Eda[9,10], José H. Garcia[1], Stephan Roche[1,11], Juan. F. Sierra[1], Sergio O. Valenzuela[1,11]\**

[1]Catalan Institute of Nanoscience and Nanotechnology (ICN2), CSIC and the Barcelona Institute of Science and Technology (BIST), Campus UAB, 08193 Bellaterra, Spain

[2]Universitat Autònoma de Barcelona, 08193 Bellaterra, Spain

[3]Institut de Ciència de Materials de Barcelona, ICMAB-CSIC, Campus UAB, 08193 Bellaterra, Spain

[4]Research Center for Electronic and Optical Materials, National Institute for Materials Science, 305-0047 Tsukuba, Japan

[5]Research Center for Materials Nanoarchitectonics, National Institute for Materials Science, 305-0044 Tsukuba, Japan

[6]Institute of Optical Materials and Technologies, Bulgarian Academy of Science, 1113 Sofia, Bulgaria

[7]Institute of Solid State Physics, Bulgarian Academy of Sciences, 1784 Sofia, Bulgaria

[8]Quantum Innovation Centre (Q. InC), Agency for Science Technology and Research (A*STAR), 138634 Singapore, Singapore

[9]Chemistry Department, National University of Singapore, 117543 Singapore, Singapore

[10]Centre for Advanced 2D Materials, National University of Singapore, 117546 Singapore, Singapore

[11]Institució Catalana de Recerca i Estudis Avançats (ICREA), 08010 Barcelona, Spain

**ABSTRACT**

Van der Waals (vdW) heterostructures enable tailored electronic and magnetic phases by stacking atomically thin layers with pristine interfaces. Here, we investigate fully 2D $Cr_2Ge_2Te_6$/$WTe_2$ heterostructures and identify a strong enhancement of ferromagnetism in $Cr_2Ge_2Te_6$ (CGT). Magnetotransport measurements across multiple devices with $WTe_2$ thicknesses ranging from monolayer to bulk reveal a robust anomalous Hall effect together with a more than twofold increase of the Curie temperature and substantially enhanced coercive fields. Interface microscopy confirms chemically abrupt vdW interfaces with no detectable interdiffusion, while control experiments rule out processing- or stray-field-induced artifacts. Our experiments and theoretical calculations demonstrate that interfacial charge transfer renders CGT conductive and that proximity-induced lattice distortions in CGT enhance exchange and magnetocrystalline anisotropy. These results establish strain-mediated lattice reconstruction as a strategy for engineering high-temperature magnetic order in 2D heterostructures and clarify that modifications within the magnetic layer itself can govern proximity effects in vdW stacks.

**KEYWORDS:** 2D Materials, Magnetism, Proximity Effects, van der Waals Heterostructures

Two-dimensional (2D) materials enable deterministic assembly of van der Waals (vdW) heterostructures with pristine interfaces, providing a powerful route to engineered electronic and magnetic phases.[1–5] In these stacks, interlayer coupling can hybridize band structures and generate proximity phenomena, while interfacial reconstruction, via charge transfer and lattice distortion, can fundamentally modify the properties of the layers. Understanding and controlling these interface-driven modifications is central to building electrically readable magnetic states and, more broadly, vdW platforms that combine magnetism with strong spin–orbit coupling materials, including quantum anomalous Hall systems.[6–9]

WTe$_2$ is a particularly appealing building block in this context. It is a semimetal whose monolayer is predicted to exhibit a quantum spin Hall phase[10] and magnetotransport experiments have reported edge-dominated transport up to 100 K.[11–13] A key challenge is identifying an insulating or semiconducting 2D magnet that can provide perpendicular magnetic anisotropy (PMA) and strong interfacial coupling, while preserving the structural and chemical integrity of the vdW stack. Known 2D ferromagnets with high Curie temperature, $T_C$, such as Fe$_3$GeTe$_2$[14] and Fe$_5$GeTe$_2$,[15] are metallic, which is a critical drawback as a conductive magnetic layer can mask the interfacial response. On the other hand, common semiconducting 2D magnets that do not contain Te, such as CrI$_3$[16] and CrBr$_3$,[17] raise another concern when paired with tellurides like WTe$_2$. Chalcogens such as Te are highly mobile and a Te-free layer can facilitate Te interdiffusion, leading to stoichiometric deviations at the interface.[18,19]

Cr$_2$Ge$_2$Te$_6$ (CGT) is therefore a compelling choice because it is a 2D ferromagnetic semiconductor with PMA down to the bilayer and it already incorporates Te in its stoichiometric lattice. As a result, CGT is less likely to act as a sink for Te from WTe$_2$, suppressing Te redistribution and stabilizing a well-defined interface. However, pristine CGT exhibits modest coercivity and a bulk $T_C^{CGT} \sim 65$ K that is substantially reduced in thin films.[20] Previous approaches to modify magnetic properties in 2D magnets have mainly relied on electrostatic gating, charge transfer or on interlayer coupling with another ferromagnet.[21–25] In CGT, electrostatic gating increases carrier density and tends to drive the magnetic anisotropy in-plane.[21,22] Charge transfer and interlayer magnetic coupling in magnetic heterostructures[23,24] offer additional routes to tune switching fields and magnetic ordering but they are fundamentally different from a fully vdW-integrated, non-magnetic/ferromagnetic heterostructure. While theoretical studies have predicted that strain or strong spin-orbit coupling proximity can increase the magnetic ordering temperature[26,27], such an increase has not been demonstrated experimentally in fully vdW-integrated platforms.[28,29]

Here, we investigate fully 2D CGT/WTe$_2$ heterostructures through magnetotransport experiments in multiple devices with WTe$_2$ thicknesses ranging from monolayer to bulk. We observe a large anomalous Hall effect (AHE) together with a pronounced strengthening of magnetism. The observed $T_C$ rises to more than 150 K, and the coercive field increases dramatically while PMA is retained. Combining interface microscopy, control experiments and first principles calculations, we show that interfacial charge transfer renders CGT conductive, and dominates the transport response, whereas proximity-induced lattice distortions in CGT are identified as the prevailing mechanism that strengthens its ferromagnetic order.

CGT and WTe$_2$ crystals were exfoliated in an inert atmosphere and vertically assembled, using a dry pick-up transfer method with hBN encapsulation, onto prepatterned Ti/Pd electrodes (see Methods in the Supporting Information for details). Five devices were fabricated and characterized, spanning WTe$_2$ thicknesses from monolayer (1L), and bilayer (2L), to trilayer (3L), few-layer (FL) and bulk. The thickness of the CGT for all devices is between 7 to 19 nm, *i.e.* all devices remain in the bulk-CGT regime. Figure 1a shows an image of the 3L device, with the CGT, WTe$_2$ and hBN crystals identified together with the prepatterned electrodes. All optical microscope images of the exfoliated flakes and completed devices are provided in Section S1 of the Supporting Information.

Figures 1b and 1c show representative Raman spectra of exfoliated CGT and WTe$_2$, and of a completed device heterostructure, respectively. The device spectra exhibit no signatures of oxidation for either material (see also Section S2). In isolated WTe$_2$ flakes, thickness can often be inferred from the relative intensities of the characteristic modes near 80, 110, and 210 cm$^{-1}$ (vertical dashed lines in Fig. 1b).[30,31] However, CGT features prominent peaks at similar Raman shifts and, in our CGT/WTe$_2$ heterostructures, the spectrum is dominated by CGT when WTe$_2$ is thin, which prevents a reliable layer count from Raman spectroscopy alone.

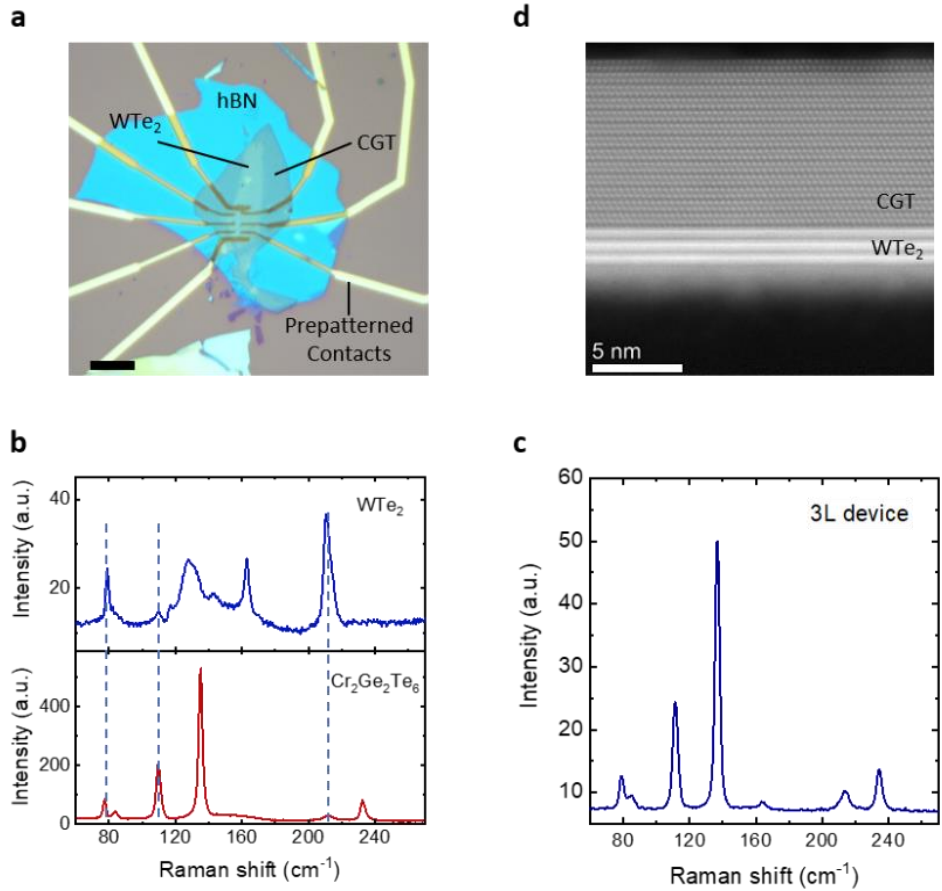

**Figure 1:** Device overview and layer identification. a) Optical microscope image of the 3L device comprising CGT and WTe$_2$ encapsulated with an hBN crystal on prepatterned Ti/Pd contacts. Scale bar is 10 μm. b) Representative Raman spectra of CGT (bottom) and WTe$_2$ (top). Vertical lines mark 80, 110 and 210 cm$^{-1}$ where thickness-sensitive WTe$_2$ Raman modes occur. c) Raman spectra obtained in the 3L device, including WTe$_2$ and CGT contributions. d) Cross-sectional HAADF-STEM image of the 3L device showing three WTe$_2$ layers.

We therefore calibrated the contrast under the optical microscope for the WTe$_2$ layer number in each device, after electrical characterization, using cross-sectional aberration-corrected scanning transmission electron microscopy in high-angle annular dark field mode (HAADF-STEM). Figure 1d shows a cross-sectional HAADF-STEM image of the 3L device, where three WTe$_2$ layers are clearly resolved (see also Section S3).

Cross-sectional HAADF-STEM imaging further reveals atomically sharp, clean CGT/WTe$_2$ interfaces without discernible structural disorder. Figure 2a shows a magnified view of the interface in the bulk-WTe$_2$ device. Lattice resolved contrast and the corresponding zone-axis alignments confirm the expected crystal structures of both WTe$_2$ and CGT. The mismatch between the zone axis orientations reflects the uncontrolled twist angle between the WTe$_2$ and CGT during heterostructure assembly.

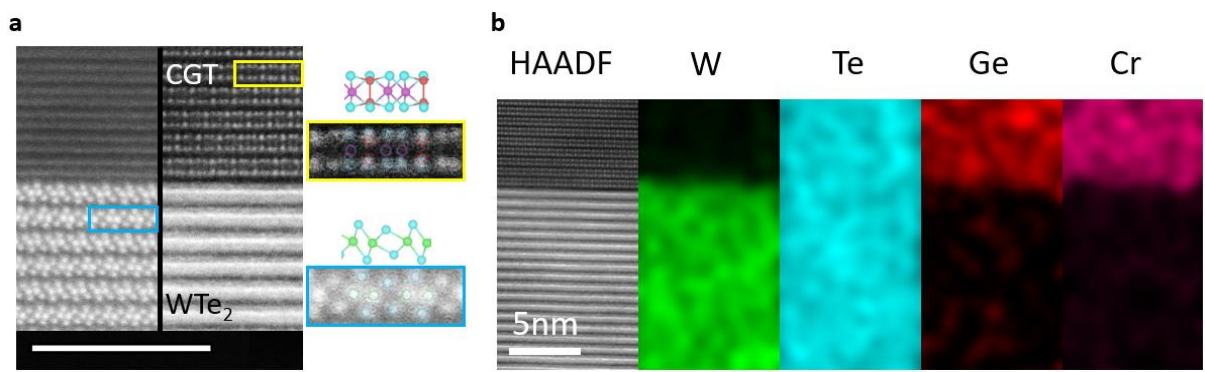

**Figure 2:** CGT/WTe$_2$ interface quality by electron microscopy. a) Representative HAADF-STEM images of the WTe$_2$ and CGT interface for the bulk device. The left image is aligned to the crystallographic [100] zone axis of WTe$_2$, while the right image is aligned to the crystallographic [100] zone axis of CGT. The crystal schematics of the regions marked with yellow (CGT) and blue (WTe$_2$) boxes were created with the software Vesta.[32] The Te, W, Ge and Cr atoms are represented with green, cyan, red and magenta spheres, respectively. Scale bar 5 nm. b) HAADF-STEM image and EDS elemental maps for the bulk device. The interface is atomically sharp with no detectable interdiffusion of W, Cr or Ge is detected within noise level.

Complementary energy-dispersive x-ray spectroscopy (EDS) elemental mapping across the interface (Fig. 2b) shows no detectable interdiffusion within the spatial resolution and sensitivity of the measurement, consistent with a clean vdW contact after fabrication. These observations hold across all devices with different $WTe_2$ thicknesses shown in Section S3 of the Supporting Information, providing strong evidence against interfacial alloying.

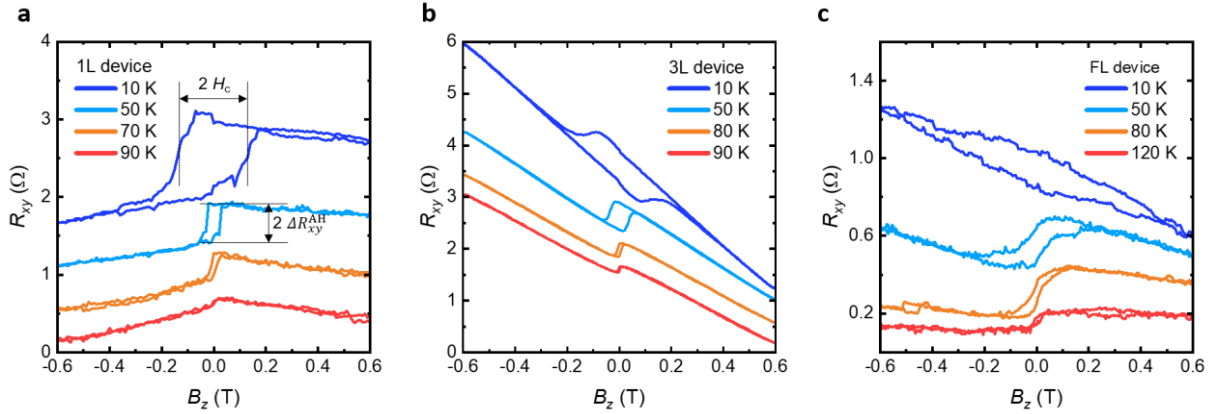

**Figure 3**: Anomalous Hall effect (AHE) in the a) 1L, b) 3L and c) FL devices. Each panel shows $R_{xy}$, as a function of out-of-plane magnetic field, $B_z$, at representative temperatures, showing hysteresis above 65 K (pristine CGT Curie temperature). Curves are offset vertically for clarity.

Having established the structural and chemical quality of the $CGT/WTe_2$ heterostructures, we turn to magnetotransport to quantify their magnetic properties. Figure 3 shows representative Hall effect measurements for the 1L, 3L, and FL devices. All devices exhibit a pronounced AHE, and the hysteresis loops are nearly square in several cases and persist to temperatures well above $T_C^{CGT}$. In the 2L and FL devices, the switching is less abrupt although the loops remain broader than those typically observed in pristine CGT (see also Section S4).[20,21,33] For the 1L device (Fig. 3a), the slanted background is attributed to an admixture of the longitudinal magnetoresistance into $R_{xy}$, due to non-ideal device geometry and slight contact misalignment. The anomalous Hall contribution is isolated by antisymmetrizing the measured transverse signal and subtracting the linear high-field ordinary Hall background. The anomalous Hall amplitude, $\Delta R_{xy}^{\text{AH}}$, is then defined as half the difference between the saturated positive and negative $R_{xy}$ branches, extrapolated to zero magnetic field.

Figure 4a compares $\Delta R_{xy}^{\text{AH}}$ across the five devices. The device Curie temperature, $T_{\text{C}}$, is obtained from the temperature at which $\Delta R_{xy}^{\text{AH}}$ vanishes. In all devices, $T_{\text{C}}$ exceeds 90 K and reaches ~155 K in the FL device, *i.e.,* more than twice $T_{\text{C}}^{\text{CGT}}$, as summarized in Table 1. The absolute magnitude of $\Delta R_{xy}^{\text{AH}}$ varies substantially between devices, mainly due to thickness-dependent current shunting through WTe$_2$ and geometric/contact variability. Consistent with this interpretation, the devices with thicker WTe$_2$ show a pronounced maximum near 75 K, while the 1L and 2L devices show an increase in $\Delta R_{xy}^{\text{AH}}$ upon cooling, as expected from the corresponding temperature dependence of the longitudinal resistance $R_{xx}$ (see Section S5). To mitigate these factors, we compute the anomalous Hall conductivity, $\sigma_{xy}^{\text{AH}}$, from $\Delta R_{xy}^{\text{AH}}$ and $R_{xx}$. Within a parallel channel description, the total conductivity tensor is the sum of the magnetic and non-magnetic contributions, with $\sigma_{xy}^{\text{AH}} \propto \frac{\Delta R_{xy}^{\text{AH}}}{R_{xx}^2}$ being largely insensitive to the non-magnetic shunt. We then normalize $\sigma_{xy}^{\text{AH}}$ to its value at $T_{\text{C}}^{\text{CGT}} \approx 65$ K, $\sigma_{xy}^{\text{AH}}(65\text{ K})$, to enable device-to-device comparison and distinguish the regimes above and below $T_{\text{C}}^{\text{CGT}}$. Figure 4b shows $\sigma_{xy}^{\text{AH}}/\sigma_{xy}^{\text{AH}}(65\text{ K})$ for all the devices. For $T > T_{\text{C}}^{\text{CGT}}$, where the bulk CGT is paramagnetic, $\sigma_{xy}^{\text{AH}}$ is dominated by the interfacial magnetic contribution. In this regime, all curves collapse onto a common decreasing trend and vanish at an enhanced $T_{\text{C}} > 90$ K. For $T < T_{\text{C}}^{\text{CGT}}$, the behavior separates into two classes: *i*) a monotonic increase upon cooling, and *ii*) a maximum near $T_{\text{C}}^{\text{CGT}}$ followed by a decrease at lower temperature.

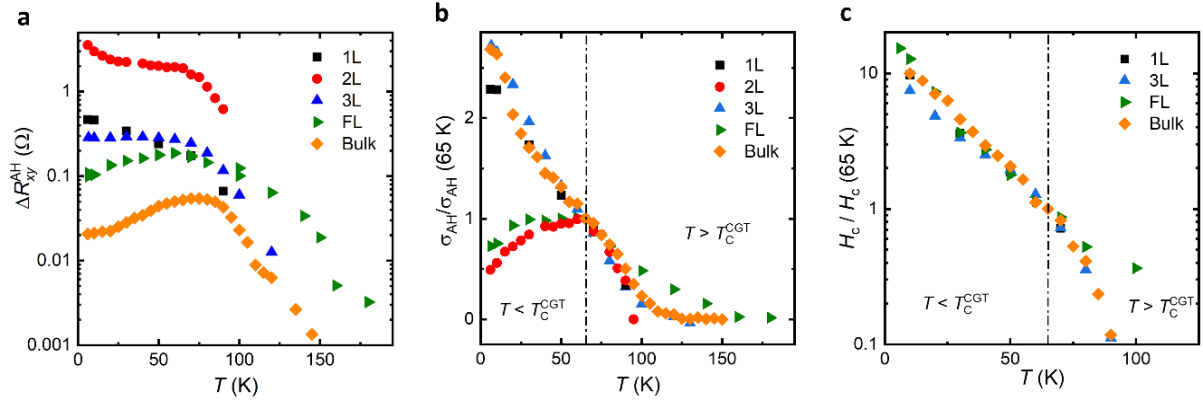

**Figure 4:** Anomalous Hall effect magnitude and coercive field. a) $\Delta R_{xy}^{AH}$ as a function of $T$ for all five devices. b) Normalized anomalous Hall conductivity $\sigma_{xy}^{AH}/\sigma_{xy}^{AH}$ (65 K). c) Normalized coercive field $H_c/H_c$(65 K) exhibiting a common exponential decay up to 65 K. $H_c/H_c$(65 K) for the 2L device not included as $H_c$(65 K) ~ 0.

The emergence of distinct low-temperature trends at $T_C^{CGT}$ supports an interfacial origin of the enhanced magnetism. Proximity to WTe$_2$ can increase the effective ordering temperature of CGT near the interface but it cannot raise the Curie temperature of the CGT bulk away from it. Therefore, $T_C^{CGT}$ remains a relevant temperature scale. A natural picture is a two-component magnetic stack comprising an interfacial region that sets the elevated $T_C$, and the bulk CGT volume that orders below $T_C^{CGT}$ and modifies the interfacial state via exchange coupling and changes in the electronic boundary conditions. For example, bulk magnetic ordering can be accompanied by shifts in the density of states, screening or chemical potential that alter the interfacial magnetization, anisotropy, and/or carrier density. Depending on how this coupling renormalizes the interface AHE contributions, $\sigma_{xy}^{AH}$ can exhibit a device-dependent temperature behavior below $T_C^{CGT}$.

| Device | $T_C$ in K | $H_c$ in mT |
|---|---|---|
| CGT | 65 | ~ 20 |
| 1L | 100 ± 10 | 135 ± 10 |
| 2L | 93 ± 3 | 46 ± 2 |
| 3L | 105 ± 5 | 135 ± 5 |
| FL | 155 ± 5 | 335 ± 30 |
| Bulk | 105 ± 3 | 170 ± 20 |

**Table 1:** Curie temperature, $T_C$, and coercive field, $H_c$, at 10 K for all five devices and pristine CGT. The $T_C$ for pristine CGT is extracted from literature,[20] while the $H_c$ was measured by graphene micromagnetometry (see Section S6).

We also quantify the coercive field, $H_c$, as the magnetic field at which the AHE loop crosses its midpoint, *i.e.*, where the AHE signal changes sign during magnetization reversal (see Fig. 3a). The resulting $H_c$ values are markedly larger than in pristine CGT, which displays little to no hysteretic behavior.[20,21,33] For comparison, the values of $H_c$(10 K) are listed in Table 1. While the absolute value of $H_c$ may be affected by extrinsic factors, such as geometry, thickness and pinning, the heterostructures show more square-like hysteresis loops, together with increased $H_c$ and $T_C$. Taken together, these trends are consistent with enhanced perpendicular magnetic anisotropy, as further supported by the DFT results discussed below.

The temperature dependence of $H_c$ summarized in Figure 4c shows that $H_c/H_c(65\,\text{K})$ decreases approximately exponentially with increasing temperature up to 65 K. The collapse of $H_c/H_c(65\,\text{K})$ across devices suggests that magnetization reversal is governed by a common thermal-activation process,[34] possibly controlled by nucleation or domain-wall depinning in the interfacial CGT region and therefore largely insensitive to device geometry. This behavior is captured by an Arrhenius-type dependence $H_c(T) \approx H_0 \exp\left(-\frac{T}{T_0}\right)$ with the effective activation scale $T_0 \approx 53$ K.

Finally, we evaluate the Hall angle $\theta_{AH}$ (defined as $\tan\theta_{AH} \approx \Delta R_{xy}^{AH}/R_{xx}$). The FL device presents the largest $\theta_{AH}$ (~0.13 deg), whereas the remaining devices cluster in the range of 0.01 - 0.03 deg despite $R_{xx}$ spanning orders of magnitude. The $\theta_{AH}$ in our devices are comparable to those reported in CGT/topological insulator heterostructures.[35–37] However, these platforms do not exhibit a substantial enhancement of $T_C$, aside from a modest ~15 K increase reported upon varying the Ge content in CGT.[38] Similarly, exfoliated devices integrated with sputtered heavy metal layers onto CGT show only small (10-20 K) increases in $T_C$.[39–41] A much larger $T_C$ increase has been reported after a high-temperature annealing (~400 °C) of CGT with a sputtered tungsten overlayer.[42] In this case, the associated anomalous Hall angle is extremely small, suggesting a different interfacial state than in our chemically clean vdW heterostructures.

To exclude that annealing or fabrication alone modifies CGT in a way that could reproduce our observations, we performed control experiments on pristine CGT processed under comparable conditions. Numerous pristine CGT flakes of varying thickness were annealed up to 400 ºC and none exhibited resistances below 150 kΩ, thereby excluding processing-induced changes in the electrical properties. We further probed the effect of annealing on CGT, and tested whether stray fields could mimic an AHE-like switching, using a graphene Hall bar micromagnetometer.[43] The high-mobility encapsulated graphene senses the stray fields from a CGT flake placed on top (see Section S6). We observe that the coercive field of the pristine CGT, as expected,[20,21,33] is an order of magnitude smaller than in CGT/WTe$_2$ devices (Table 1) and remains unchanged after annealing. Moreover, the stray fields are too weak to cause the observed AHE-like switching, as reproducing the measured Hall amplitudes would require stray fields of ~50 mT (the bulk device) up to ~4 T (the 1L device). These controls rule out processing-and stray-field-driven artifacts, supporting an interfacial origin of the observed AHE and enhanced coercivity in CGT/WTe$_2$.

Collectively, our transport and control experiments provide strong evidence against a scenario in which the enhanced $T_C$ originates solely from proximity-induced magnetism in WTe$_2$. Instead, they point to a modification of CGT at the interface, potentially coexisting with proximity effects in WTe$_2$. First, induced moments in a proximitized semimetal are typically weak and are unlikely to account for the observed enhancement without an accompanying change in CGT, as quantitatively assessed in the numerical analysis below. In addition, the $T_C$ does not scale with the AHE amplitude, suggesting changes in intrinsic magnetic energy scales, with no systematic correlation to current partitioning between the layers. Moreover, interfacial charge transfer that increases the CGT conductivity can account for the large AHE. However, a doping-only mechanism, analogous to ionic gating, would soften the perpendicular anisotropy or reorient it in-plane[22] (as confirmed by our first-principles calculations; see below), and is thus incompatible with the robust PMA observed in our devices. HAADF-STEM/EDS further show no detectable intermixing and the $T_C$ enhancement does not appear with just annealing, arguing against chemical alloying or changes in bulk CGT, respectively.

Together, these constraints require a mechanism that *i*) strengthens the magnetic energy scales, *ii*) preserves PMA and chemical integrity of the materials, and *iii*) originates specifically at the CGT/WTe$_2$ interface, and not just by charge transfer. This points to proximity-induced lattice distortion of the interfacial CGT, stabilized by WTe$_2$, which enhances exchange and magnetocrystalline anisotropy and thereby boosts both $T_C$ and $H_c$ well beyond those of pristine CGT. In the calculations that follow, we show that realistic CGT distortions in the heterostructure reproduce the observed $T_C$ increase, and the coercivity enhancement, while preserving the PMA, whereas charge transfer alone from WTe$_2$ cannot achieve this.

To rationalize the origin of the enhanced magnetic ordering, we performed density functional theory (DFT) calculations (see computational details in Methods) on both the pristine materials and the CGT/WTe$_2$ heterostructure. Electronic band structure simulations (see Section S7) reveal a metallic ground state in CGT induced by interfacial hybridization with WTe$_2$ in the CGT/WTe$_2$ heterostructure. A direct comparison between the heterostructure and the pristine materials highlights significant band hybridization, consistent with the reduced interlayer distance ($d \sim 2$ Å) obtained from our structure optimizations.

Bader charge analysis and band alignment calculations were performed to quantify the extent of charge transfer at the interface. While a significant electron transfer from WTe$_2$ to CGT of $\sim 2 \times 10^{13}$ e$^-$/cm$^2$ is observed, driven by the Fermi-level realignment, spin density simulations indicate a negligible spin transfer between the two materials (see Section S8). These results identify charge transfer as the origin of the enhanced conductivity of the CGT interface, while the magnetism remains localized in CGT, consistent with an interface modified magnetic layer rather than magnetism hosted by WTe$_2$.

Additionally, structural optimization of the WTe$_2$/CGT heterostructure reveals that the heterostructure environment induces an inhomogeneous biaxial distortion in the CGT layer. This distortion is characterized by unequal expansion along the two in-plane directions, consistent with the orthorhombic symmetry of WTe$_2$. The relaxed structure yields an average in-plane expansion of about 3%. HAADF-STEM analysis reveals a smaller, but qualitatively consistent, tensile lattice modulation in CGT near the interface, confirming that the distortion is predominantly located in the CGT layer (Section S3 of the Supporting Information).

We therefore divided our simulations into four stages: *i*) pristine, *ii*) doped, and *iii*) distorted CGT, as well as *iv*) the CGT/WTe$_2$ heterostructure, to systematically analyze the electronic, structural and magnetic properties of these materials and to disentangle the effects of electron transfer and strain.

To investigate the magnetic properties, we combined DFT + U calculations with Green's function methods and model the materials using a Heisenberg spin Hamiltonian. Atomistic simulations were subsequently carried out to construct large-scale samples and estimate the critical temperatures. Our calculations successfully reproduce the experimental findings, yielding a $T_C$ >150 K in the heterostructure, as shown in Figure 5a.

Our combined DFT, spin-model, and atomistic simulations also indicate that the enhancement of $T_C$ originates primarily from WTe$_2$-induced strain in the CGT lattice, which strongly modifies the leading exchange parameters (Fig. 5b). Structural distortions that break inversion symmetry also activate non-negligible Dzyaloshinskii–Moriya interactions (DMI). However, due to the dominant increase in isotropic exchange, we do not expect DMI to play a major role in determining $T_C$. Exchange parameters and DMI defined according to the spin Hamiltonian in Eq. S1, are reported in Tables S2 and S3 in Section S9 of the Supporting Information.

Magnetic anisotropy energy (MAE) calculations further reveal that both the heterostructure and distorted CGT exhibit triaxial magnetic anisotropy, in contrast to pristine CGT, which displays a well-defined out-of-plane easy axis and in-plane hard-plane. Previous studies have shown that electron doping in pristine CGT can reorient the magnetization into the plane.[22] While our calculations reproduce this trend, they also show that the non-pristine systems recover and even strengthen the out-of-plane anisotropy due to strain-induced symmetry breaking. Specifically, we obtain a MAE of 131 μeV/Cr for pristine CGT, which increases to 304 μeV/Cr in the heterostructure. In other words, strain enhances the MAE to a level sufficient to preserve the out-of-plane easy axis despite the electron doping introduced by interfacial charge transfer.

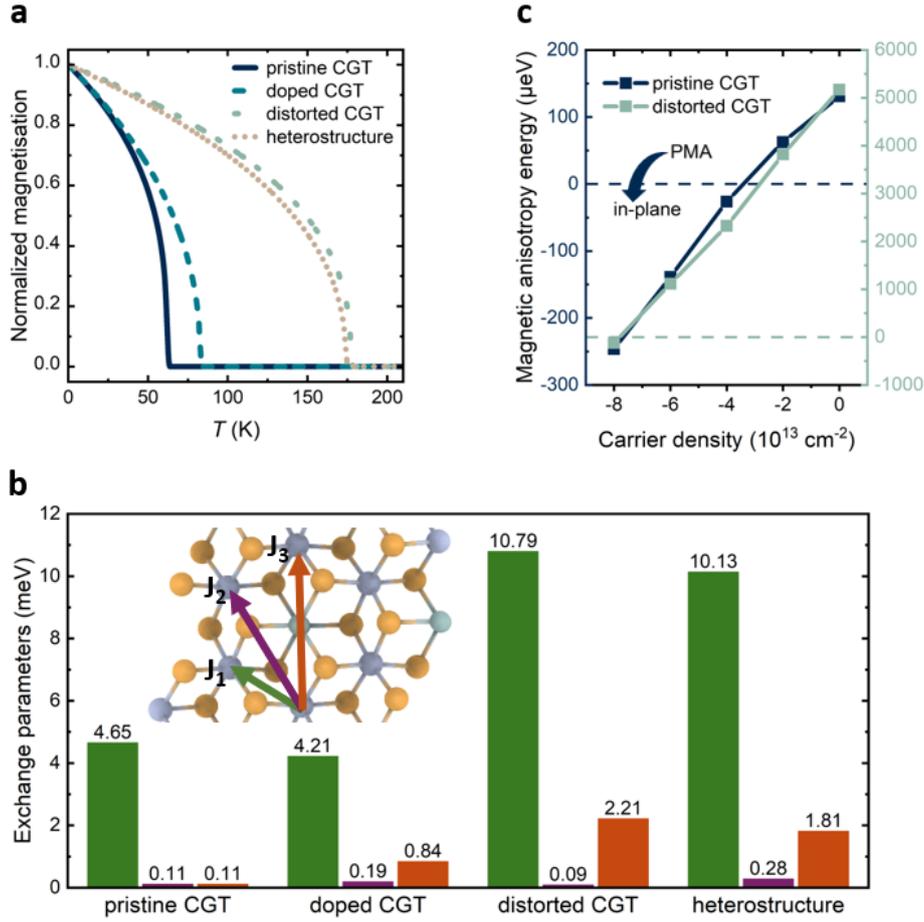

**Figure 5:** Calculated magnetic properties. a) CGT magnetization calculations, showing the changes in $T_C$, for pristine CGT (solid line) and after including the effect of doping (dashed lines) or lattice distortion (dot/double-dot line). The dotted line represents the calculated magnetization when both doping and distortion are included, as expected in the CGT/WTe$_2$ heterostructure. b) Exchange parameters for the interaction between nearest neighbors, $J_1$, second-nearest neighbors, $J_2$, and third-nearest neighbors, $J_3$, for the pristine, doped and distorted CGT, and the CGT/WTe$_2$ heterostructure. c) The magnetic anisotropy energy as a function of carrier density varied by gating or charge transfer. A transition from perpendicular to in-plane magnetic anisotropy can be observed for both pristine and distorted CGT, however, at different densities.

Figure 5c shows the MAE of pristine CGT and of an isolated distorted CGT layer as a function of electron doping. Here, the distorted CGT is an auxiliary model to isolate the role of structural distortion. It is obtained by extracting the CGT geometry from the relaxed $WTe_2$/CGT heterostructure, while removing the $WTe_2$ layer. The simulations reproduce the previously reported tendency towards in-plane magnetism in pristine CGT with hole doping but show that this tendency is suppressed in distorted CGT. In this case, structural distortions strongly modify the orbital hybridization and reconstructs the band structure, leading to a markedly enhanced MAE. However, this model is not intended to reproduce the MAE in the full $WTe_2$/CGT heterostructure, for which the MAE is determined by the interplay of distortion and interfacial interactions.

To complete our analysis, we performed controlled doping simulations to disentangle the effect of charge transfer on both exchange interactions and $T_C$. These calculations confirm that charge transfer plays only a secondary role, in contrast to the strong enhancement of exchange interactions and $T_C$ induced by strain. Importantly, when modelling pristine CGT with the combined effects of structural distortions and electron doping, the results fully reproduce those of the heterostructure. This demonstrates that the phenomenon can be reduced to these two ingredients, which together provide an excellent match to the heterostructure simulations. Figures 5 shows the evolution of the main exchange parameters and critical temperatures across the four cases considered: pristine CGT, CGT with the level of electron doping estimated from the charge transfer, isolated distorted CGT, and the heterostructure. The results clearly show that strain is the dominant factor in reproducing the heterostructure behavior, while electron doping provides a modest compensating contribution, leading to quantitative agreement with the heterostructure.

In conclusion, we demonstrated that vdW CGT/$WTe_2$ heterostructures exhibit a dramatically enhanced magnetic order, with Curie temperatures exceeding 150 K and coercive fields far surpassing those of pristine CGT. By combining detailed magnetotransport analysis, interface microscopy, and first-principles calculations, we reveal a mechanism that fundamentally differs from the commonly assumed proximity-induced exchange in WTe$_2$. Instead, interfacial charge transfer renders CGT conductive, while proximity-induced lattice distortions emerge as the dominant driver of the enhanced ferromagnetic order and magnetic anisotropy in CGT. These results highlight that often-overlooked structural degrees of freedom in vdW stacks can be engineered to control magnetic energy scales. By establishing strain-mediated proximity as a reliable route to enhancing magnetism, this study paves the way for high-temperature, electrically addressable magnetic interfaces in next-generation quantum and spintronic devices.

## ASSOCIATED CONTENT

Supporting Information: Methods for device fabrication, electrical characterization, STEM measurements and computational details. Raman spectra and thickness determination. Additional HAADF-STEM images and analysis, EDS maps and magnetotransport data. Micromagnetometry measurements. Calculated band structure and alignment, charge and spin transfers, exchange and DMI parameters.

## AUTHOUR INFORMATION

**Corresponding Author**

*corresponding author: franz.herling@icn2.cat; mireia.torres@icn2.cat; SOV@icrea.cat

**Authors' contribution**

F.H().[†] and M.T.-S.[†] fabricated the samples and performed the measurements and data analysis. M.A. and C.E. performed the sample fabrication, measurements and data analysis for the micro-magnetometry study. D.L.E. performed the theoretical calculations under the supervision of J.H.G. and S.R. K.X. and J.S.R. characterized the Raman modes of the devices. M.R. performed TEM lamella fabrication, and K.G. and B.M. the TEM measurements. K.W. and T.T. grew the h-BN crystals, I.V. and G.E. the CGT crystals, and D.D. and V.M. the $WTe_2$ crystals. J.F.S. supervised the day-to-day laboratory work of M.T.-S. and C.E., in coordination with S.O.V. S.O.V. conceived and supervised the research and provided overall scientific guidance. All authors contributed to the manuscript preparation or commented on the manuscript.

[†]These authors contributed equally

**ACKNOWLEDGEMENTS**

We thank M. Tena for preliminary transport studies on CGT/WTe$_2$ heterostructures. This research was partially supported by the HEDOS project (PID2022-143162OB-I00) funded by MICIU/AEI/10.13039/501100011033 and FEDER and by the Severo Ochoa Program CEX2021-001214-S. The authors acknowledge the ICN2 Research Support Division for providing access to laboratory facilities and technical assistance and the use of instrumentation as well as the technical advice provided by the Joint Electron Microscopy Center at ALBA (JEMCA) and funding from Grant IU16-014206 (METCAM-FIB) to ICN2 funded by the European Union through the European Regional Development Fund (ERDF), with the support of the Ministry of Research and Universities, Generalitat de Catalunya. F.H. acknowledges funding through Grant JDC2023-052262-I funded by MICIU/AEI/10.13039/501100011033 and by the European Social Fund Plus (ESF+). M.T.-S. acknowledges support by the Ajuts Joan Oró grant for the recruitment of predoctoral research staff in training (2025 FI-1 00113), funded by the Department of Research and Universities of the Government of Catalonia, and co-funded by the ESF+. D.L.E., J.G. and S.R. acknowledge funding from MCIN/AEI/10.13039/501100011033, under grants PID2019-106684GB-I00 and PID2022-138283NB-I00 and by ERDF, and ERC project AI4SPIN funded by Horizon Europe-European Research Council Executive Agency under grant agreement No 101078370; MICIU with European funds-NextGenerationEU (PRTR-C17.I1); and 2021 SGR 00997 funded by Generalitat de Catalunya. They also acknowledge the use of HPC resources at MareNostrum and the technical support provided by Barcelona Supercomputing Center (RES-FI-2025-3-0058). K.X. and J.S.R. acknowledge financial support by MICIU/AEI/10.13039/501100011033 under grants PDC2023-145934-I00 and PID2024-162811NB-I00 and by the Severo Ochoa Centres of Excellence Program under grant CEX2023-001263-S. V. M. and D. D. acknowledge funding from the Bulgarian National

Science Fund (BNSF) under project No. KP-06-H 98/7. G.E. acknowledges support from the Ministry of Education (MOE), Singapore, under AcRF Tier 1 grant (A-8001995-00-00).**REFERENCES**

(1) Geim, A. K.; Grigorieva, I. V. Van Der Waals Heterostructures. *Nature* **2013**, *499* (7459), 419–425. https://doi.org/10.1038/nature12385.
(2) Novoselov, K. S.; Mishchenko, A.; Carvalho, A.; Neto, A. H. C. 2D Materials and van Der Waals Heterostructures. *Science* **2016**, *353* (6298), aac9439. https://doi.org/10.1126/science.aac9439.
(3) Žutić, I.; Matos-Abiague, A.; Scharf, B.; Dery, H.; Belashchenko, K. Proximitized Materials. *Mater. Today* **2018**. https://doi.org/10.1016/j.mattod.2018.05.003.
(4) Sierra, J. F.; Fabian, J.; Kawakami, R. K.; Roche, S.; Valenzuela, S. O. Van Der Waals Heterostructures for Spintronics and Opto-Spintronics. *Nat. Nanotechnol.* **2021**, *16*, 856–868. https://doi.org/10.1038/s41565-021-00936-x.
(5) Frisenda, R.; Navarro-Moratalla, E.; Gant, P.; Pérez De Lara, D.; Jarillo-Herrero, P.; Gorbachev, R. V.; Castellanos-Gomez, A. Recent Progress in the Assembly of Nanodevices and van Der Waals Heterostructures by Deterministic Placement of 2D Materials. *Chem. Soc. Rev.* **2018**, *47* (1), 53–68. https://doi.org/10.1039/C7CS00556C.
(6) Hellman, F.; Hoffmann, A.; Tserkovnyak, Y.; Beach, G. S. D.; Fullerton, E. E.; Leighton, C.; MacDonald, A. H.; Ralph, D. C.; Arena, D. A.; Dürr, H. A.; Fischer, P.; Grollier, J.; Heremans, J. P.; Jungwirth, T.; Kimel, A. V.; Koopmans, B.; Krivorotov, I. N.; May, S. J.; Petford-Long, A. K.; Rondinelli, J. M.; Samarth, N.; Schuller, I. K.; Slavin, A. N.; Stiles, M. D.; Tchernyshyov, O.; Thiaville, A.; Zink, B. L. Interface-Induced Phenomena in Magnetism. *Rev. Mod. Phys.* **2017**, *89* (2). https://doi.org/10.1103/RevModPhys.89.025006.
(7) Mak, K. F.; Shan, J.; Ralph, D. C. Probing and Controlling Magnetic States in 2D Layered Magnetic Materials. *Nat. Rev. Phys.* **2019**, *1* (11), 646–661. https://doi.org/10.1038/s42254-019-0110-y.
(8) Tokura, Y.; Yasuda, K.; Tsukazaki, A. Magnetic Topological Insulators. *Nat. Rev. Phys.* **2019**, *1* (2), 126–143. https://doi.org/10.1038/s42254-018-0011-5.
(9) Choi, E.-M.; Sim, K. I.; Burch, K. S.; Lee, Y. H. Emergent Multifunctional Magnetic Proximity in van Der Waals Layered Heterostructures. *Adv. Sci.* **2022**, *9* (21), 2200186. https://doi.org/10.1002/advs.202200186.
(10) Qian, X.; Liu, J.; Fu, L.; Li, J. Quantum Spin Hall Effect in Two-Dimensional Transition Metal Dichalcogenides. *Science* **2014**, *346* (6215), 1344–1347. https://doi.org/10.1126/science.1256815.
(11) Fei, Z.; Palomaki, T.; Wu, S.; Zhao, W.; Cai, X.; Sun, B.; Nguyen, P.; Finney, J.; Xu, X.; Cobden, D. H. Edge Conduction in Monolayer $WTe_2$. *Nat. Phys.* **2017**, *13* (7), 677–682. https://doi.org/10.1038/nphys4091.
(12) Tang, S.; Zhang, C.; Wong, D.; Pedramrazi, Z.; Tsai, H.-Z.; Jia, C.; Moritz, B.; Claassen, M.; Ryu, H.; Kahn, S.; Jiang, J.; Yan, H.; Hashimoto, M.; Lu, D.; Moore, R. G.; Hwang, C.-C.; Hwang, C.; Hussain, Z.; Chen, Y.; Ugeda, M. M.; Liu, Z.; Xie, X.; Devereaux, T. P.; Crommie, M. F.; Mo, S.-K.; Shen, Z.-X. Quantum Spin Hall State in Monolayer 1T'-$WTe_2$. *Nat. Phys.* **2017**, *13* (7), 683–687. https://doi.org/10.1038/nphys4174.
20

**GRAPHICAL TOC**

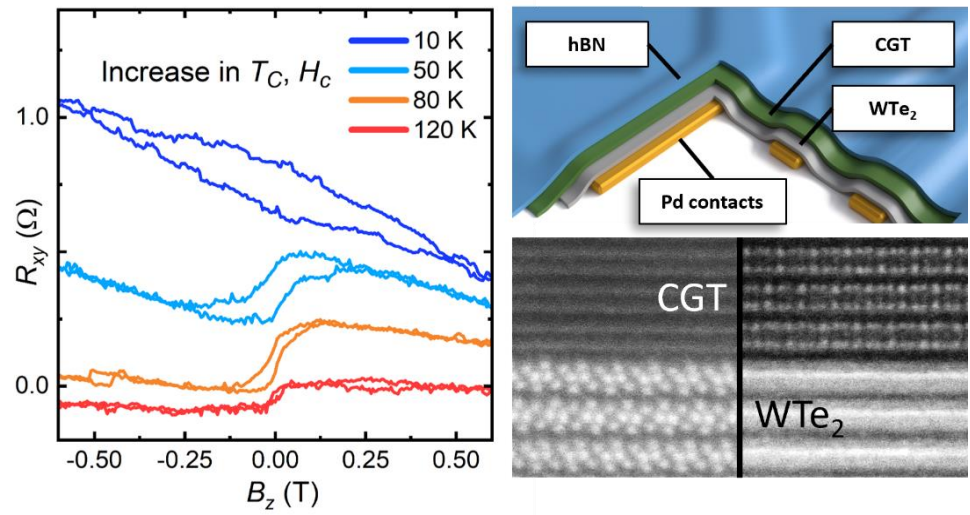

*Supplementary Information*

# Strain-Mediated Lattice Reconstruction Enhances Ferromagnetism in Cr$_2$Ge$_2$Te$_6$/WTe$_2$ van der Waals Heterobilayers

*Franz Herling*[1†*], *Mireia Torres-Sala*[1,2†*], *Dorye. L. Esteras*[1], *Charlotte Evason*[1], *Motomi Aoki*[1], *Marcos Rosado*[1], *Kapil Gupta*[1], *Bernat Mundet*[1], *Kai Xu*[3], *J. Sebastián Reparaz*[3], *Kenji Watanabe*[4], *Takashi Taniguchi*[5], *Dimitre Dimitrov*[6,7], *Vera Marinova*[6], *Ivan A. Verzhbitskiy*[8], *Goki Eda*[9,10], *José H. Garcia*[1], *Stephan Roche*[1,11], *Juan. F. Sierra*[1], *Sergio O. Valenzuela*[1,11]*

[1]Catalan Institute of Nanoscience and Nanotechnology (ICN2), CSIC and the Barcelona Institute of Science and Technology (BIST), Campus UAB, 08193 Bellaterra, Spain

[2]Universitat Autònoma de Barcelona, 08193 Bellaterra, Spain

[3]Institut de Ciència de Materials de Barcelona, ICMAB-CSIC, Campus UAB, 08193 Bellaterra, Spain

[4]Research Center for Electronic and Optical Materials, National Institute for Materials Science, 305-0047 Tsukuba, Japan

[5]Research Center for Materials Nanoarchitectonics, National Institute for Materials Science, 305-0044 Tsukuba, Japan

[6]Institute of Optical Materials and Technologies, Bulgarian Academy of Science, 1113 Sofia, Bulgaria

[7]Institute of Solid State Physics, Bulgarian Academy of Sciences, 1784 Sofia, Bulgaria

[8]Quantum Innovation Centre (Q. InC), Agency for Science Technology and Research (A*STAR), 138634 Singapore, Singapore

[9]Chemistry Department, National University of Singapore, 117543 Singapore, Singapore

[10]Centre for Advanced 2D Materials, National University of Singapore, 117546 Singapore, Singapore

[11]Institució Catalana de Recerca i Estudis Avançats (ICREA), 08010 Barcelona, Spain

**CONTENTS**

## S1 METHODS

**Device fabrication**

The WTe$_2$ crystals were grown by chemical vapor transport method as described in Ref. [1]. The CGT growth followed the procedure described in Ref. [2]. Details of the hBN growth can be found in Ref. [3]. The WTe$_2$, CGT and hBN flakes were mechanically exfoliated inside a glovebox with an Ar atmosphere onto p-doped Si/SiO$_2$ substrates. Thin flakes of WTe$_2$ were identified by optical contrast, calibrated with scanning transmission electron microscopy (STEM), and categorized as monolayer, bi-, tri-, or few-layer. WTe$_2$-CGT-hBN vdW heterostructures were fabricated using a dry stamping technique.[4] For the assembly, a suitable hBN flake was initially picked up using a viscoelastic stamp made of polydimethyl-siloxane/polycarbonate. The hBN layer was then used to sequentially pick up the CGT and the WTe$_2$ layers via vdW forces, before transferring the complete stack onto prepatterned contacts by melting the stamp at 180 ºC and dissolving any residues with chloroform, acetone and isopropanol. The prepatterned Ti/Pd electrodes were defined using electron-beam lithography with resist masks fabricated via spin-coating methyl methacrylate and polymethyl methacrylate and deposited via electron-beam evaporation in an ultra-high vacuum chamber ($10^{-8}$ torr). The assembled were annealed at 280°C (320 °C for the 2L device) for 3 hours in a high-vacuum chamber ($10^{-7}$ Torr) to improve the CGT/WTe$_2$ interface quality.

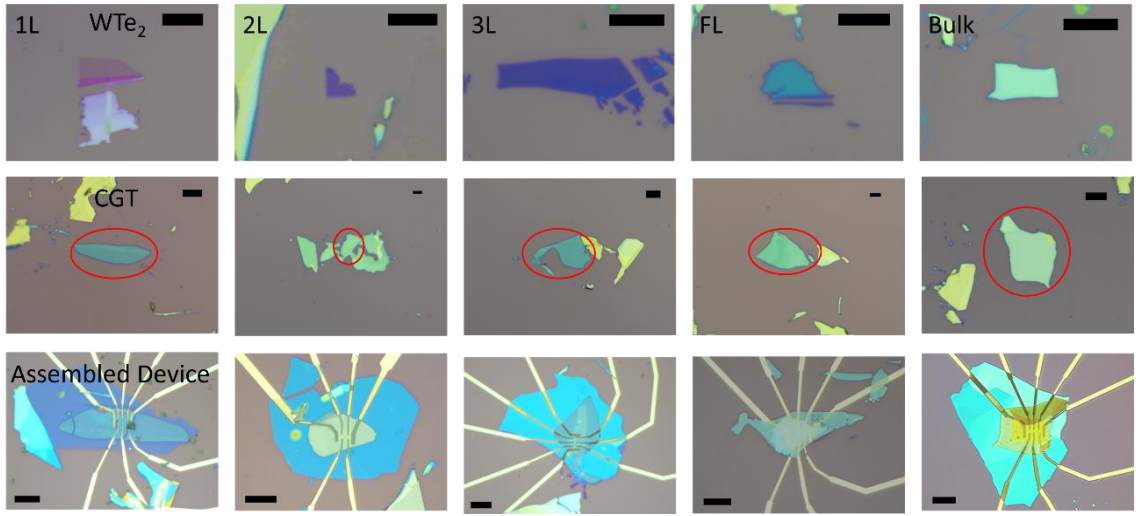

**Figure S1:** Optical microscope images of exfoliated WTe$_2$ (top) and CGT (middle) flakes before pick up and transfer onto the pre-patterned Ti/Pd contacts. The assembled devices are shown in the bottom panel. From left to right: 1L, 2L, 3L, FL, and bulk. Scale bars 10 μm

**Electrical characterization**

For the magneto-transport measurements, the devices were placed in a cryo-free cryostat. An electrical current of 10 or 100 μA was supplied using a Keithley 6221 current source operating with a Keithley 2182A nanovoltmeter in current reversal (delta) mode, or a Stanford Research Systems 810 lock-in at an alternating current (AC) frequency of $f_{AC}$ = 177 Hz to measure the voltage.

**Scanning transmission electron microscopy measurements**

An electron transparent STEM lamella was prepared in a Dual Beam Helios 5 UX Focused ion beam. Atomic-resolution high-angle annular dark-field (HAADF) STEM images were acquired in a double-corrected Thermofisher SPECTRA 300 (S)TEM microscope operated at either 300KV or 200KV. For the HAADF images, a convergence and collection semi-angles of around 20mrad and 67-200mrad were used respectively. Energy-dispersive x-ray spectroscopy (EDS) spectrum images were acquired using a window-less 4 quadrant SuperX detector and the compositional maps were generated following the Cliff-Lorimer methodology.

**Computational details**

DFT + U calculations were performed using the Siesta package[5] with a Hubbard U parameter of 1 eV for the Cr atoms. The choice of U was validated by simulating all relevant properties across the range 1–5 eV in steps of 0.5 eV, with a finer refinement of 0.1 eV near the optimal values. The results proved robust over the entire range, but values of U = 1-3 eV yielded electronic band gaps and exchange parameters in closer agreement with literature.[6,7] The exchange–correlation energy was treated within the generalized gradient approximation using the Perdew-Burke-Ernzerhof[8] functional and standard pseudopotentials from the PseudoDojo database[9], with spin–orbit coupling explicitly included. All the calculations were performed with a real-space mesh cutoff of 1200 Ry. The Brillouin zone was sampled using a Monkhorst–Pack mesh[10] of 12 × 12× 1 for pristine CGT and 8 × 8 × 2 for the heterostructure. Van der Waals interactions were accounted for using Grimme-D3 semi-empirical corrections.[11] Exchange interactions were obtained via Green's function calculations implemented in the TB2J package[12], with careful convergence tests on the simulation cell size. Critical temperatures were estimated using the Vampire[13] software with cell dimensions of 50 nm ×50 nm ×1 nm equilibrated over 40000 steps and averaged over 60000 steps. The heterostructure geometry was obtained through a fully self-consistent structural optimization. The distorted CGT layer was subsequently extracted from the relaxed structure after removing the $WTe_2$ layer. All references to strain and structural features arise from this self-consistent relaxation and are not from externally imposed distortions.

## S2 THICKNESS DETERMINATION OF THE WTE$_2$ FLAKES

To determine the thicknesses of WTe$_2$ flakes, the intensities of the 80, 110 and 210 cm$^{-1}$ Raman peaks have been used.[14,15] However, as indicated by the red arrows in Figure S2a and b, CGT has peaks at very similar Raman shifts which makes an analysis difficult.

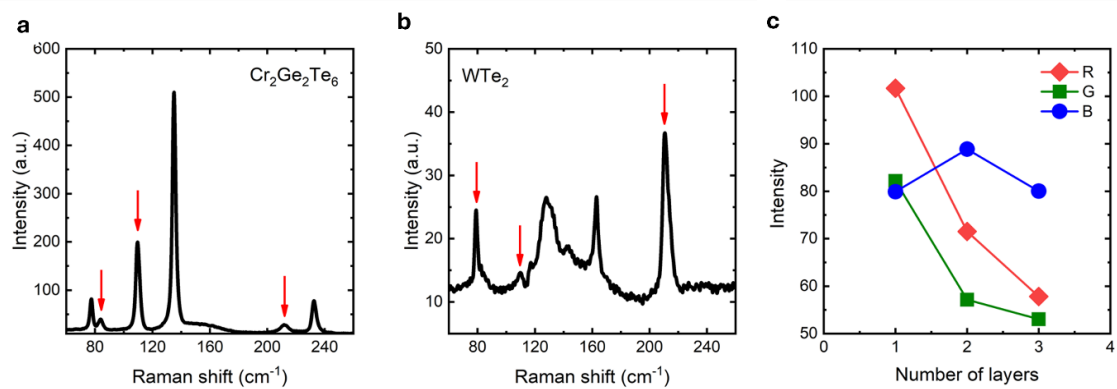

**Figure S2:** Raman spectra and thickness determination. a) Raman spectrum of a CGT and b) WTe$_2$ flake. c) Intensity of the red, green and blue channel for WTe$_2$ flakes with different thicknesses (from Fig. S1).

Therefore, we used the intensity of the different color channels of the optical microscope images, calibrated with the STEM image of the trilayer sample, to determine the number of layers. As shown in Figure S2c, especially the red and green channel can be used to precisely distinguish between the flakes.

Figure S3 shows the Raman spectra of the CGT/WTe$_2$ bilayer for the 1L, 2L and 3L device (same as Fig 1c in the main text). It should be noted that none of the Ramen spectra recorded for all of the devices show an indication for oxidation of the CGT (no double peak around 80 cm$^{-1}$) or WTe$_2$ flakes (for example a WO$_x$ peak at 270 cm$^{-1}$).[16]

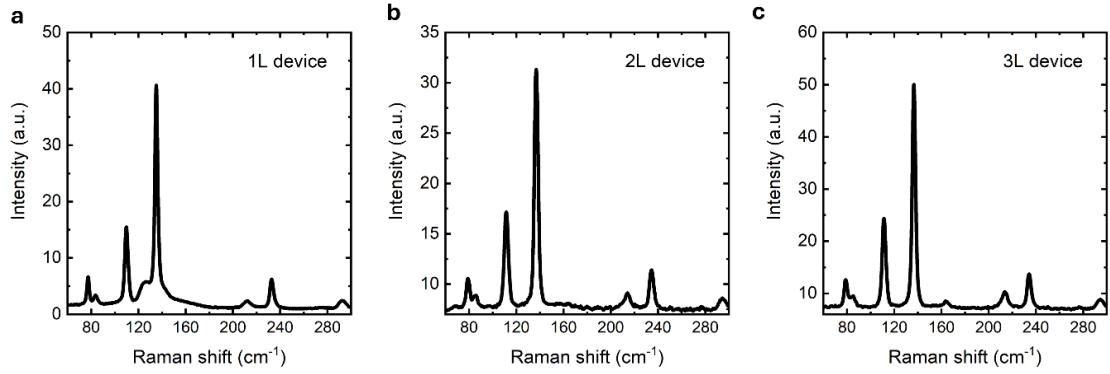

**Figure S3:** Raman spectra of the CGT/WTe$_2$ bilayer. Measured for the a) 1L, b) 2L and c) 3L device.

Additionally, the angle-dependent polarized Raman spectra can be used to determine the crystal orientation of the material.[18] Figure S4b presents these data for the 1L device and as expected the crystal axes are parallel to the cleaving edges of the exfoliated flakes.

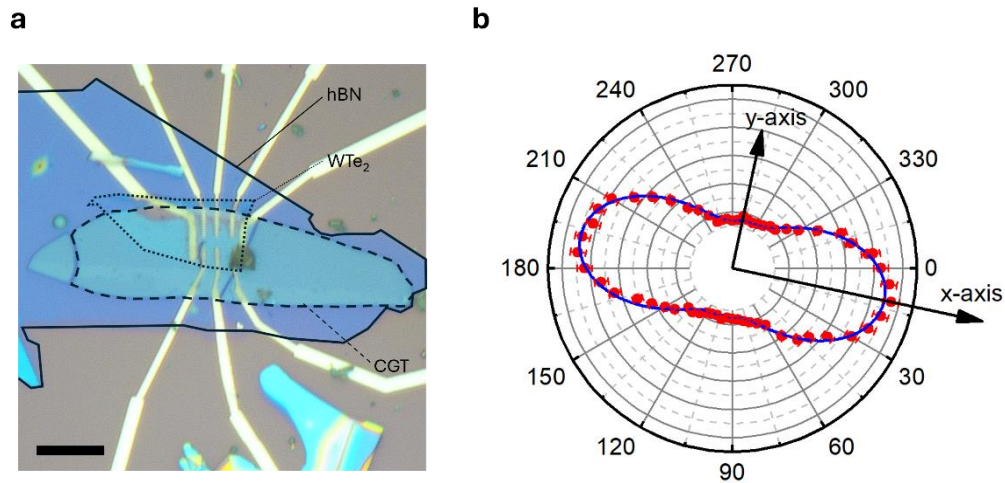

**Figure S4:** Characterization of the crystal axis of the WTe$_2$ flakes. a) Optical microscope image of the 1L device. Scale bar is 10 μm. b) Angular dependence of the polarized Raman spectra of the Ag mode at 212.7 cm$^{-1}$ to identify the crystalline axes of the device in a. Red scatter plot is measured intensity values with the error bars representing the standard deviation of the mean value, the blue line is a fit to the data. The crystal axes are similar to the cleaving edges of the flake.

## S3 HIGH-ANGLE ANNULAR DARK FIELD SCANNING TRANSMISSION ELECTRON MICROSCOPY (HAADF-STEM) IMAGES AND ENERGY-DISPERSIVE X-RAY SPECTROSCOPY (EDS) COMPOSITIONAL MAPS

To study the interface between $WTe_2$ and CGT after sample fabrication and electrical measurements, a cross-sectional, electron transparent STEM lamella was prepared by focused ion beam. Atomic-resolution high-angle annular dark-field (HAADF) STEM images and EDS spectrum images were acquired, and compositional maps generated (see Methods). Figure S5 shows all HAADF STEM images of the five devices.

While the pristine crystal structure of the CGT is visible for the 1L device, we were not able to image the monolayer $WTe_2$ below. This could be due to the preparation of the lamella as this was the thinnest lamella of the five devices. For the 2L device, the lamella was prepared on the edge of the bilayer $WTe_2$ flake, across one of the prepatterned contacts, and possibly due to this, it was again not possible to image the $WTe_2$. The FL device is the only device that was not encapsulated with hBN so the CGT flake is more deteriorated after the lamella preparation process than in the other devices. Therefore, only for the 3L and bulk devices could the twist angle between the layers be determined, as explained in Figure S6. While the difference between the two crystallographic directions is very similar for the two devices, the resulting twist angle between the two layers is different.

The STEM-HAADF image and EDS maps for the 2L device are shown in Figure S7. As expected from the STEM-HAADF images in Figure S5, no W was detected, but as the lamella was prepared on the edge of the flake through one of the prepattern contacts, the absence of diffusion of the Pd from the metal electrodes can be observed. Around the electrodes, the encapsulation of the hBN is not perfect due to step in height, so small traces of oxidation can be seen.

For the 3L, FL and bulk device in Figure S8, the $WTe_2$ and CGT layer can be observed clearly, with no sign of oxidation and diffusion of the four elements.

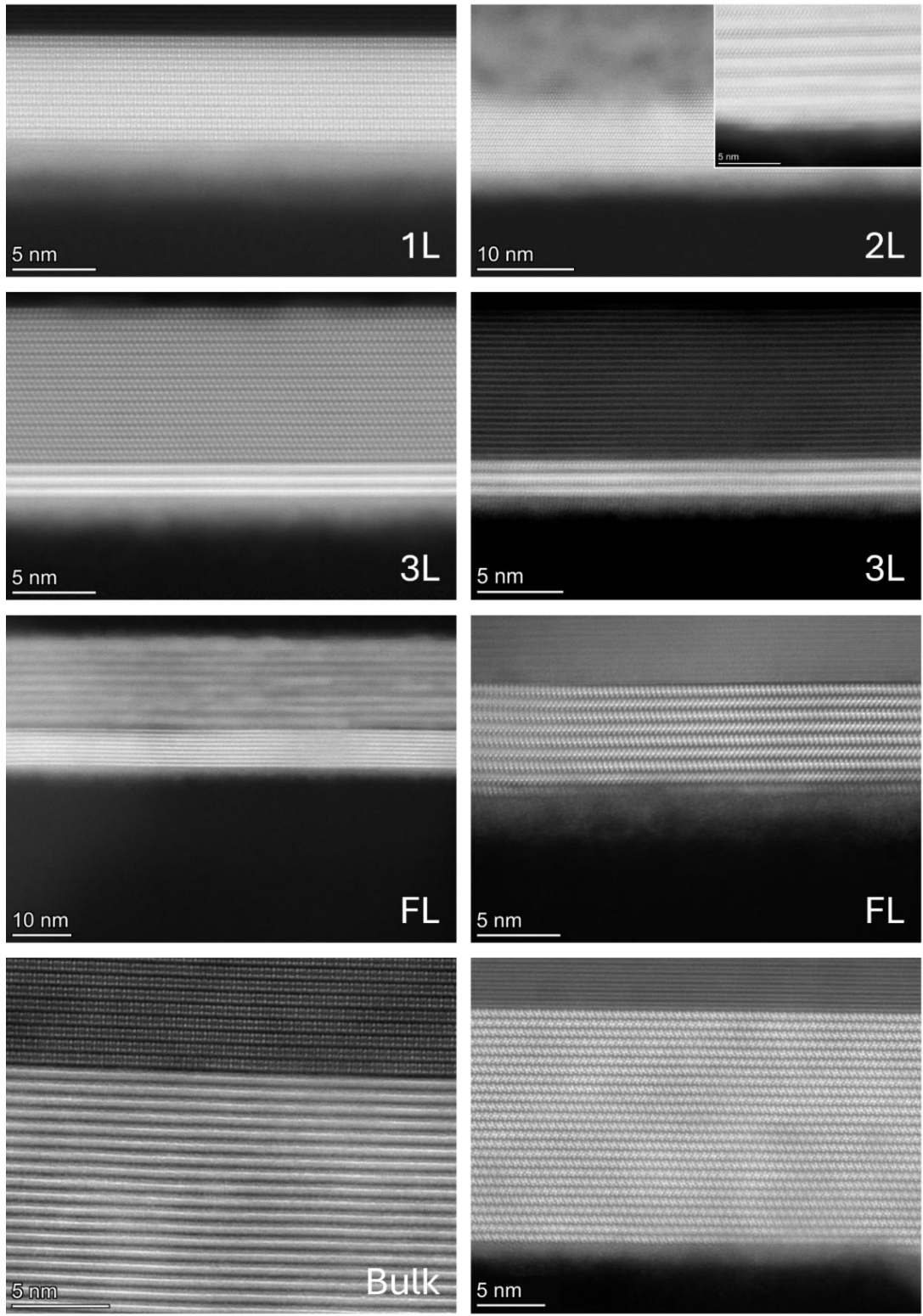

**Figure S5:** STEM-HAADF images of all 5 devices. For the 2L device a zoomed in image is shown as inset. For the 3L, FL and bulk device, the left image is aligned to the crystallographic [100] zone axis of CGT, while the right image is aligned to the crystallographic [100] zone axis of $WTe_2$. The difference between the two zone axes is between 12-15 deg for each device.

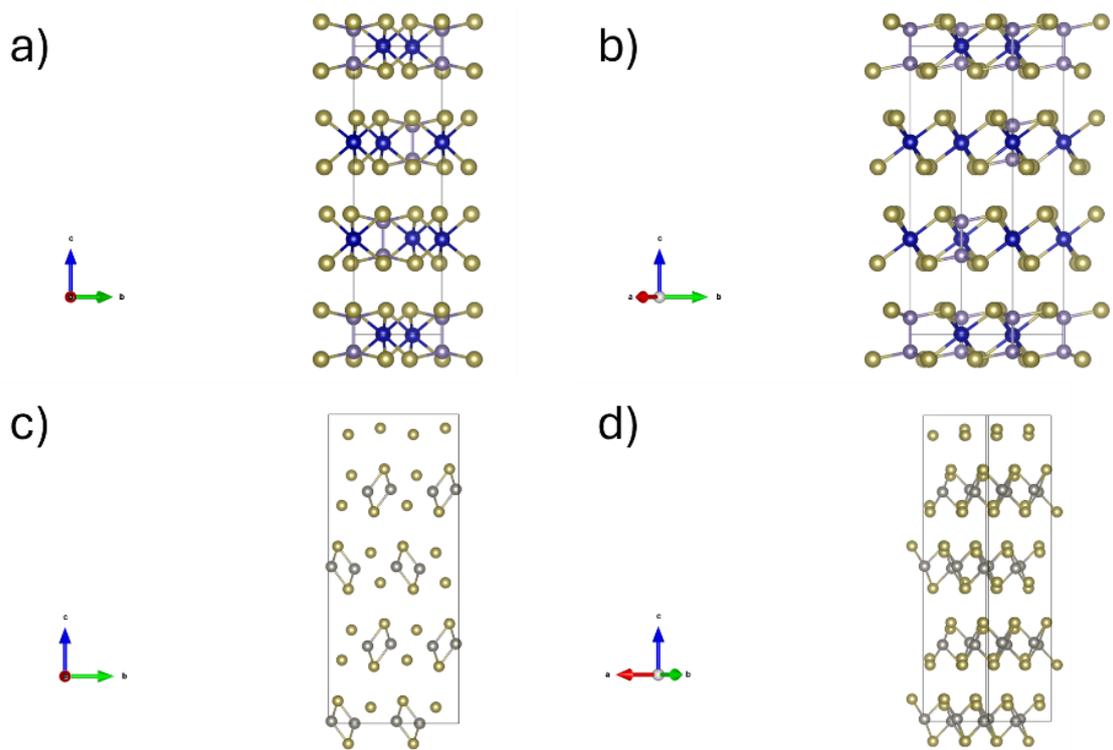

**Figure S6:** Crystal orientation of the STEM-HAADF images shown above, created with Vesta.[17] a) CGT, shown along the [100] zone axis as for the 1L and bulk device and b) oriented along the [001] zone axis as for the 2L and 3L device. c) WTe$_2$, shown along the [100] zone axis as for the 3L and FL device and d) oriented along the [001] zone axis as for the bulk device.

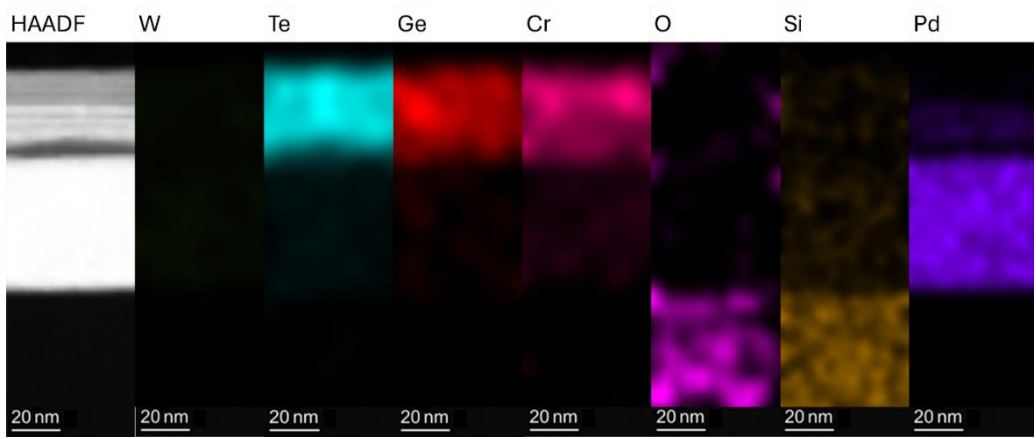

**Figure S7:** STEM-HAADF image and EDS elemental maps for the 2L device.

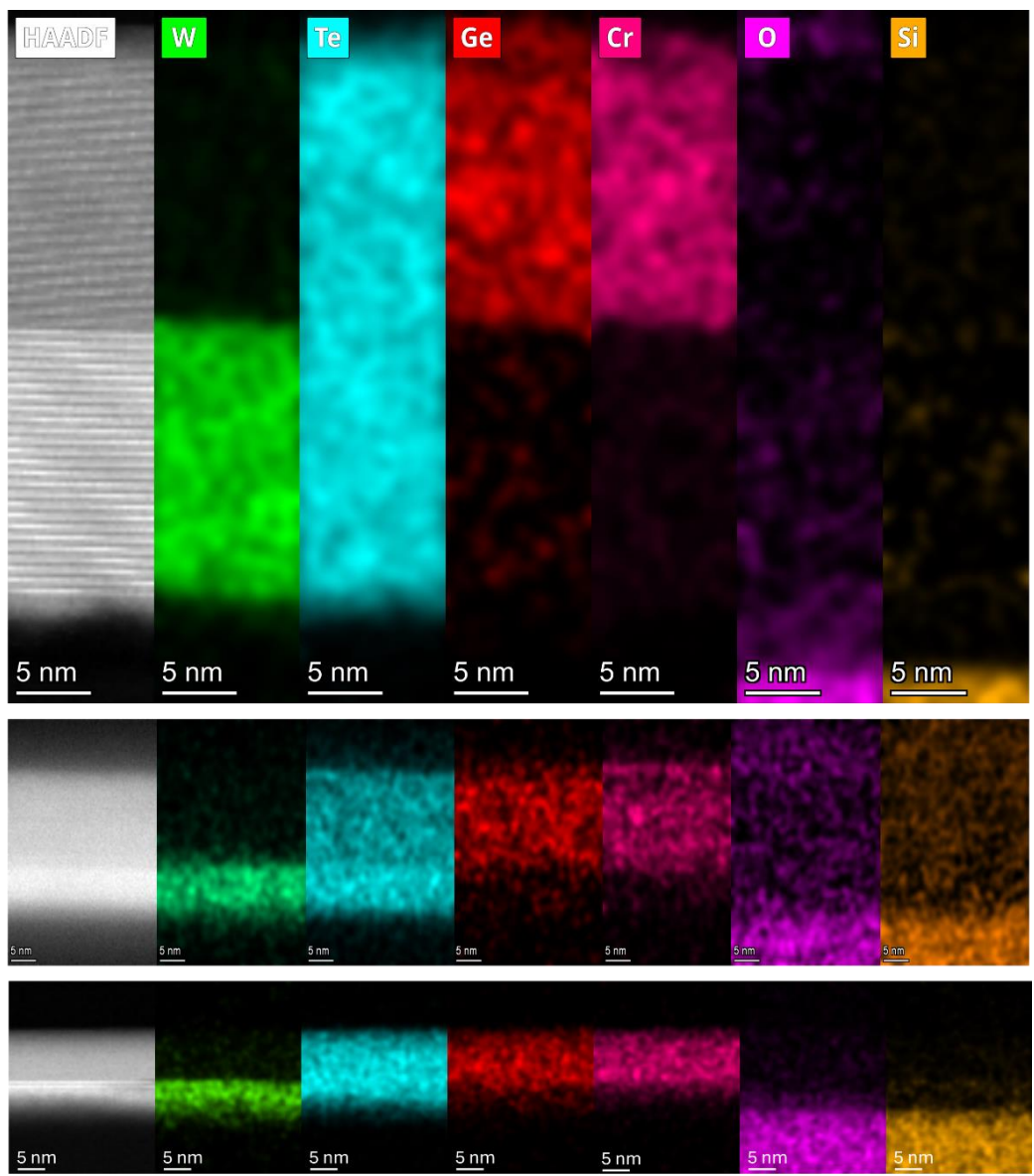

**Figure S8:** STEM-HAADF image and EDS elemental maps for the bulk, Fl and 3L device.

The HAADF-STEM image of the bulk device was used to quantify the change in the lattice of CGT. The central position of the Te columns was identified using a 2D gaussian profile fitting using the ATOMAP documentations[19] and then averaged the in-plane interspacing over 24 atomic columns. The relative change when going away from the interface is plotted in Figure S9b.

For the in-plane modulation, a tensile strain of around 0.8 % can be extracted when comparing the first layer at the interface with WTe$_2$ with the one being 42 layers away.

However, even at these distances from the interface no full relaxation of the strain can be observed yet. It should be noted that the effect of the hBN capping layer on top could already play a role in these layers as it is only another 3 nm away.

This analysis was also performed for the bulk $WTe_2$ layer and no in-plane modulation could be observed.

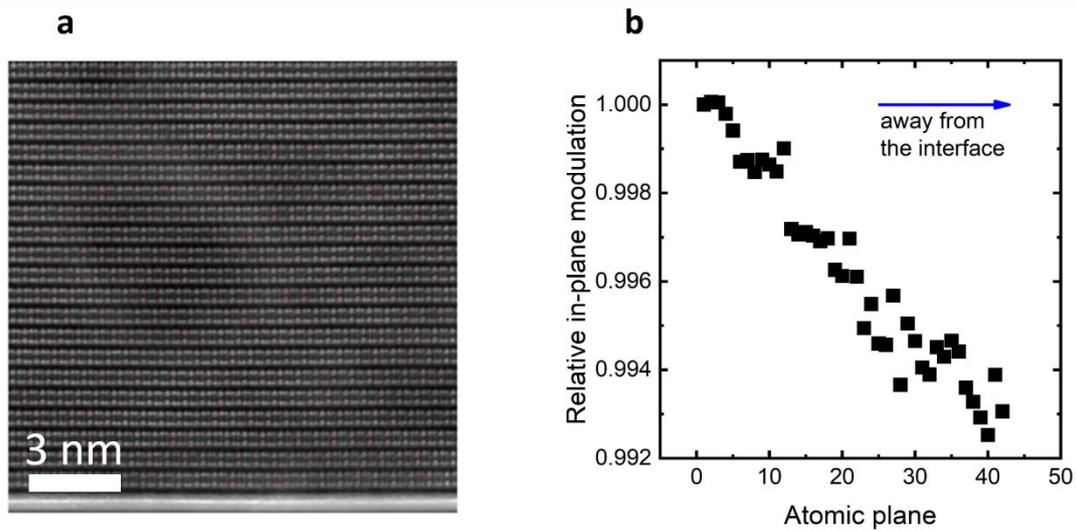

**Figure S9:** Analysis of the STEM image for the bulk device. a) HAADF-STEM images with the position of the Te columns marked by red dots. b) Relative in-plane modulation, of the CGT layer from panel a plotted as a function of the number of atomic planes away from the interface.

## S4 ANOMALOUS HALL EFFECT MEASUREMENTS

Corresponding to Figure 3 in the main text, we show here the Hall effect measurements for the 2L and bulk devices.

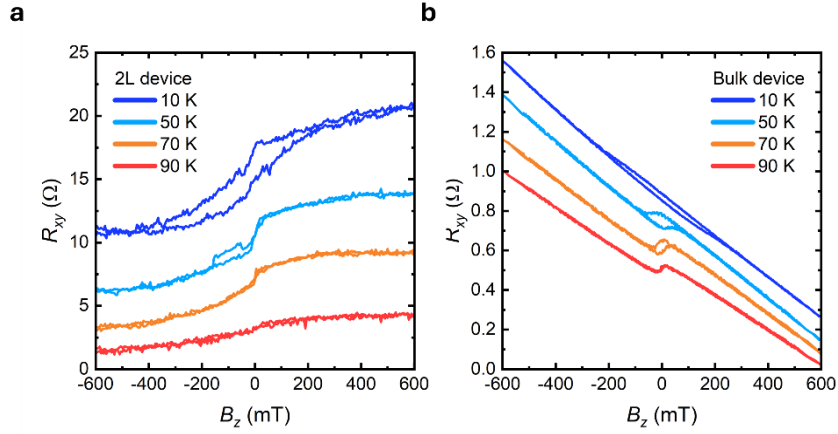

**Figure S10:** Anomalous Hall effect (AHE) measurements for the a) 2L and b) bulk device, completing the measurement data shown in the main text. Each panel shows $R_{xy}$, as a function of out-of-plane magnetic field, $B_z$, at representative temperatures. Curves offset for clarity. The small opening in panel a) at 50 K is most likely due to a measurement artefact, as no comparable feature is observed at the neighboring temperatures.

## S5 TEMPERATURE DEPENDENCE OF THE ANOMALOUS HALL ANGLE AND THE LONGITUDINAL RESISTANCE

As mentioned in the main text, we define the anomalous Hall angle $\theta_{AH}$ via $\tan\theta_{AH} \approx \Delta R_{xy}^{AH}/R_{xx}$. Figure 11a shows the temperature dependence of $\tan\theta_{AH}$. Notably, as $H_c$, $\tan\theta_{AH}$ tracks the enhanced $T_C$, consistent with a strengthening of intrinsic magnetic energy scales. The FL device presents the largest $\theta_{AH}$ (~0.13 deg) with the largest magnetic ordering temperature, whereas the remaining devices cluster in the range of 0.01 - 0.03 deg despite $R_{xx}$ spanning orders of magnitude.

Complementary, Figure S11b shows the temperature dependence of the longitudinal four-probe resistance, $R_{xx}$, of the five devices. The values are normalized by the 10 K value to facilitate comparison of the trends. While the bulk device exhibits strong temperature dependence, this behavior is reduced for thinner devices. The 1L device shows almost constant resistance across the whole temperature range. However, this is not a monotonic trend, the FL device shows less resistance variation than the 3L sample and the 2L device has opposite temperature dependence. The temperature dependence indicates that the number of WTe$_2$ layers is not the only parameter that varies between the devices as is inherent to exfoliated and transferred 2D devices. Twist angle, strain, quality of the vdW interface and alignment of the crystal axis with the current direction could also play a role.

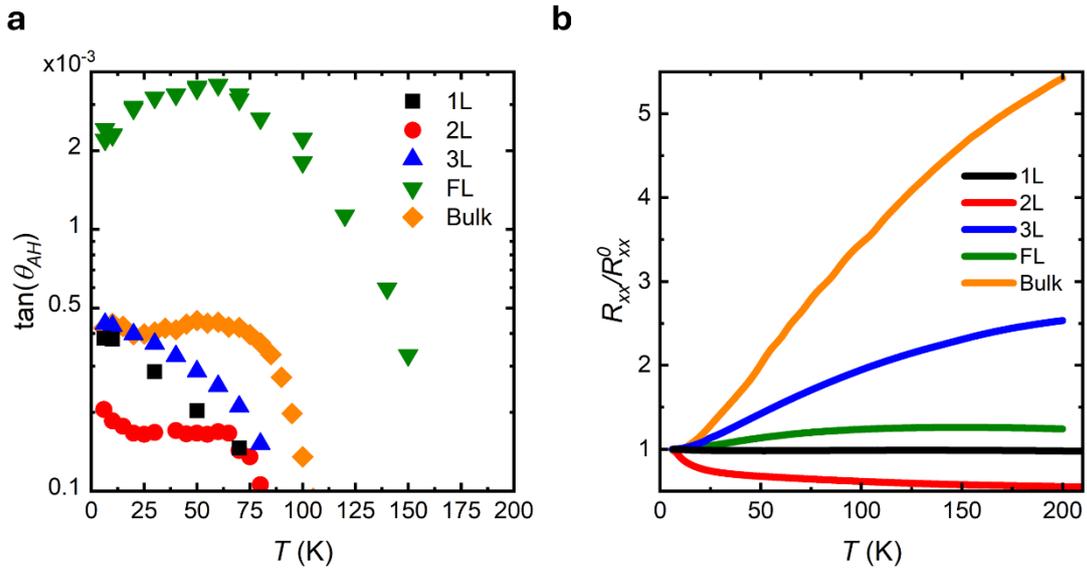

**Figure S11:** Temperature dependence of a) anomalous Hall angle and b) longitudinal resistance, $R_{xx}$, as a function of temperature, $T$, normalized by the resistance at 10 K, $R_{xx}^0$, for all five devices.

## S6 MICROMAGNETOMETRY

To test the effect of annealing on the magnetism of CGT, we wanted to directly sense the stray fields from an exfoliated flake overcoming any problems due to electrical contacting and degradation. As the volume of 2D magnets is too small for conventional magnetometry methods, we used the effect of the stray fields on the Hall curve of a conductor directly underneath the flake. They add an additional perpendicular magnetic flux through the active area of the sensor – in this case the Hall cross. Due to the hysteresis in the switching of magnetization, the difference in up and down magnetic field sweeps shows the coercivity of the CGT flake.

We therefore fabricated a fully encapsulated, high-mobility graphene Hall bar micromagnetometer as introduced in Ref. [20] to sense the stray fields from a CGT flake placed on top. The fabrication steps are as follows:

1. $O_2$ plasma cleaning and baking of the $Si/SiO_2$ (285 nm) substrates.
2. Mechanical exfoliation of h-BN (Taniguchi and Watanabe) and graphene (Natural graphite, NGS) flakes.
3. Assembly of the hBN/monolayer graphene/hBN stack with a polydimethyl-siloxane/polycarbonate stamp and dry transfer onto the substrate.
4. Annealing in high vacuum ($10^{-7}$ torr) at 350 °C for 3 h.
5. Electron-beam lithography (EBL) for designing the electrodes.
6. Reactive ion etching (RIE) with $CHF_3$ (40 sccm) + $O_2$ (4 sccm) for 3 min to make one-dimensional contacts.
7. Electron-beam deposition of Cr (5 nm)/Au (50 nm) for the electrodes in an ultra-high vacuum chamber ($10^{-8}$ torr).
8. EBL for designing the Hall bar.
9. RIE with $CHF_3$ (40 sccm) + O2 (4 sccm) for 3 min to shape the stack into a Hall bar.

10. Exfoliation of CGT and assembly of a hBN/CGT/hBN heterostructure.

11. Dry transfer of the hBN/CGT/hBN stack onto the Hall bar.

12. Annealing in high vacuum ($10^{-7}$ torr) at 280, 330 and 400 °C for 3 h (ramp rate 2 K/min).

Before and after every annealing step (12.), we measured the Hall resistance in a perpendicular magnetic field in the configuration shown in the inset in Figure S12a. The results are shown in the same panel. When subtracting the up and down field sweep as in Figure S12b, we can observe distinct switches at around ±20 mT that can be understood as the coercive fields of the CGT flake. Consistent with the $T_C$ of pristine CGT, the switches disappear at temperatures above 65 K.

We observe that the coercive field of the pristine CGT, as expected,[21] is an order of magnitude smaller than in our CGT/WTe$_2$ devices and remains unchanged even after annealing steps at higher temperatures than used during device fabrication.

Moreover, we can calculate the stray fields that would be necessary to cause the observed AHE-like switching seen in our CGT/WTe$_2$ devices $B_{stray} = \Delta R_{xy}^{\text{AH}}/R_{xy}^{H}$, by using the slope of the measured $R_{xy}$ curves, $R_{xy}^{H}$, and assuming that the amplitude $\Delta R_{xy}^{\text{AH}}$ stems fully from the additional magnetic flux.

The required stray fields range from ~50 mT (the bulk device) up to ~4 T (the 1L device) and are larger than what we observe from the CGT flake, therefore, ruling out processing-and stray-field-driven artifacts, supporting an interfacial origin of the observed AHE and enhanced coercivity in CGT/WTe$_2$.

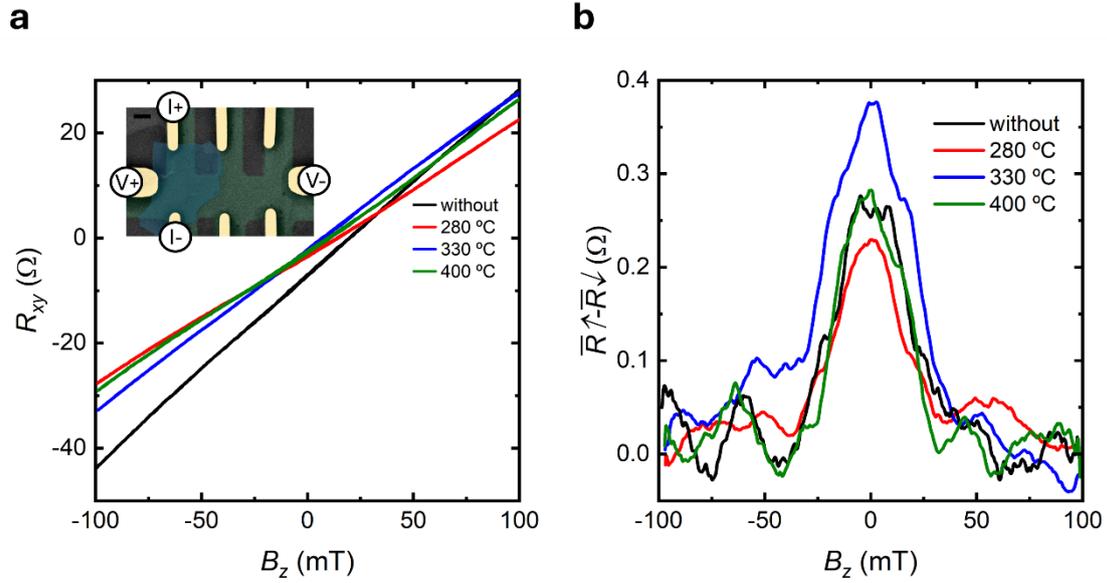

**Figure S12:** Micromagnetometry study of a pristine CGT flake. a) Hall resistance at 10 K measured before and after different annealing steps at Vg = 26 V. Inset left: False-colored scanning electron microscope image of the device, showing the contacts (yellow), graphene fully encapsulated in hBN (green) and CGT on top (blue) as well as the contacts used for the measurement of the Hall resistance. b) Difference of the averaged up and down sweep of the curves in the main graph. The switches with similar amplitude at around ±20 mT are visible in all four curves.

## S7 ELECTRONIC BAND STRUCTURE

We show here the results of the electronic band structure simulations following the workflow described in the computational details section of the Methods. The direct comparison in Figure S13 between the two pristine materials and the heterostructure highlights significant band hybridization, consistent with the reduced interlayer distance ($d \sim 2$ Å) obtained from our structure optimizations. They also reveal a metallic ground state in CGT induced by interfacial hybridization with $WTe_2$ in the $CGT/WTe_2$ heterostructure.

We observe an inhomogeneous in-plane expansion as well. This distortion is difficult to characterize due to its non-uniform nature; however, it results in an asymmetric biaxial stretching of the lattice, consistent with the orthorhombic symmetry of $WTe_2$. We obtain expansions of 2.29% along the a axis and 4.45% along the b axis, corresponding to an average strain of approximately 3.37%. A bond-length analysis shows that the Cr–Cr distances increase within a range of 2.6–4%, while the Cr–Te bonds are compressed by about 3%.

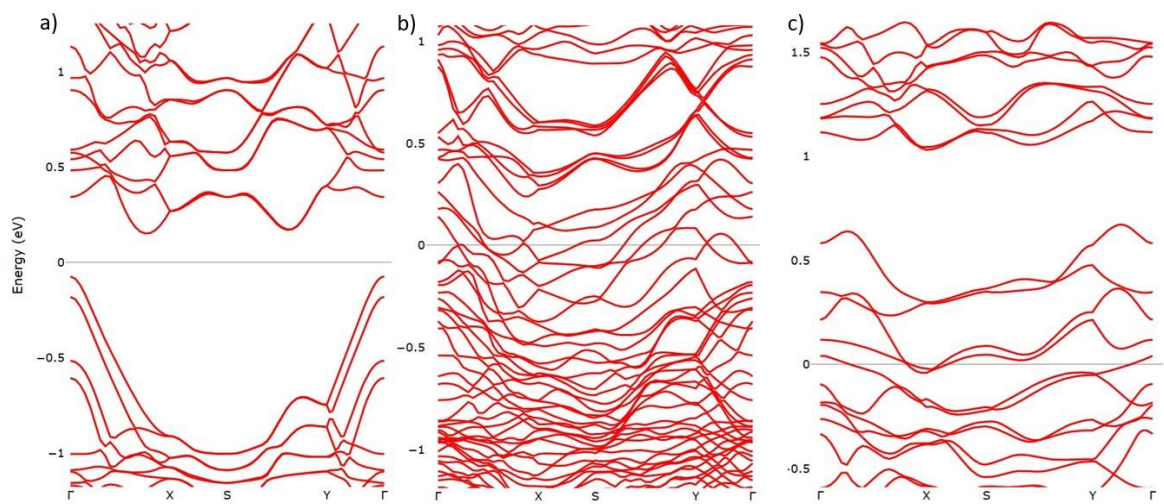

**Figure S13:** Calculated band structure of a) pristine CGT, b) $CGT/WTe_2$ heterostructure c) CGT distorted by the proximity to $WTe_2$. The band structure is calculated along the high symmetry points of the rectangular CGT cell used for the construction of the heterostructure.

**S8 CHARGE AND SPIN TRANSFER ANALYSIS**

Bader charge analysis and band alignment calculations were performed to quantify the extent of charge transfer at the interface (see Fig. S14a) due to the resulting work functions (see Fig. S15). While a significant electron transfer from $WTe_2$ to CGT of $\sim 2 \times 10^{13}$ $e^-/cm^2$ is observed (all results of the quantitative analysis can be found in Table S1), driven by the Fermi-level realignment, spin density simulations indicate a negligible spin transfer between the two materials as shown in Figure S14b.

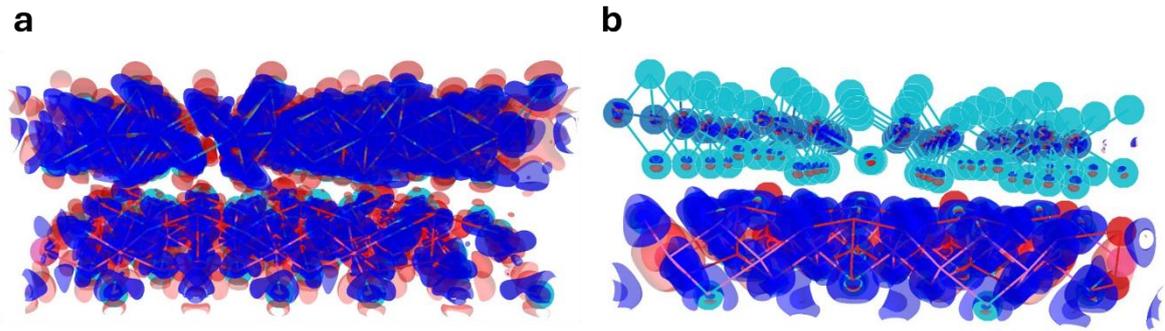

**Figure S14:** Calculated a) charge and b) spin transfer in the CGT/WTe$_2$ heterostructure. Red (blue) densities represent the gain (depletion) of charge or spin density, indicating an important transfer along the interface between both materials for charge and a spin density, that is mostly contained within CGT.

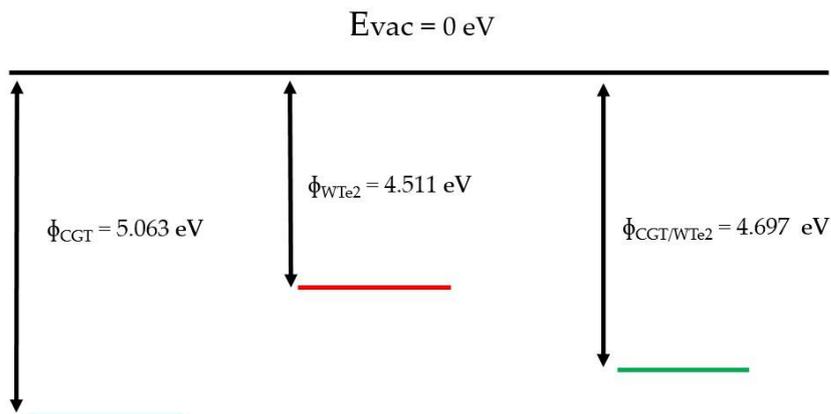

**Figure S15:** Band alignment of the CGT/WTe$_2$ heterostructure. CGT and WTe$_2$ work functions are calculated from the crystal structures in the heterostructure environment, showcasing the transference of electrons from WTe$_2$ to CGT.

**Table S1:** Quantitative Bader analysis of charge transfer. Negative numbers refer to accumulation of negative charge (electrons).

|  | Total charge in CGT | Total charge in WTe$_2$ |
|---|---|---|
| e$^-$/ cell | -0.189 | 0.186 |
| e$^-$/ cm$^2$ | -2.086 × 10$^{13}$ | 2.053 × 10$^{13}$ |

**S9 SPIN HAMILTONIAN**

The following spin Hamiltonian was used to describe the exchange in the heterostructure.

**Equation S1:** Spin Hamiltonian definition, following TB2J convention.

$$H = -\sum_{i \neq j} J_{ij}^{iso} (\vec{S}_i \cdot \vec{S}_j) - \sum_{i \neq j} \vec{D}_{ij} \cdot (\vec{S}_i \times \vec{S}_j) - \sum_{i \neq j} \vec{S}_i \mathbf{J}_{ij}^{ani} \vec{S}_j - \sum_i \vec{S}_i \mathbf{K}_i \vec{S}_i$$

Following this definition, Tables S2 and S3 show the calculated exchange parameters and Dzyaloshinskii–Moriya interactions (DMI) for the nearest, second-nearest and third-nearest neighbors.

**Table S2:** Exchange parameters in meV.

| Neighbor | CGT | Distorted CGT | Heterostructure |
|---|---|---|---|
| 1 | 4.656 | 10.790 | 10.131 |
| 2 | 0.111 | 0.093 | 0.275 |
| 3 | 0.115 | 2.213 | 1.808 |

**Table S3:** DMI modulus in meV.

| Neighbor | CGT | Distorted CGT | Heterostructure |
|---|---|---|---|
| 1 | 0.002 | 0.882 | 0.345 |
| 2 | 0.110 | 0.673 | 0.399 |
| 3 | 0.001 | 0.670 | 0.801 |

Complementary to the datasets shown in Figure 5b and c in the main text, we plot the doping-dependence of the exchange parameters in Figure S16.

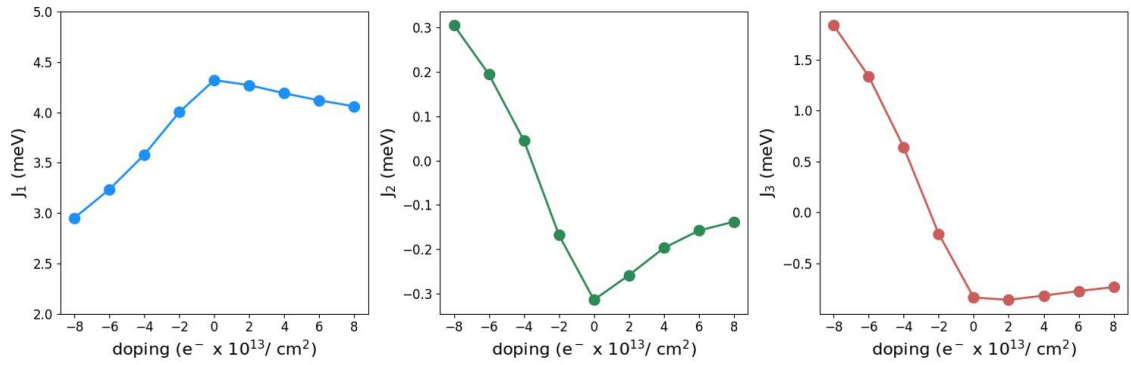

**Figure S16:** Evolution of exchange parameters of CGT under doping conditions. Negative sign represents hole doping.